\documentclass{article}
\usepackage{hiph-art}
\edyear{2006}                             
\def\rightmark{Triple-Gluon and Triple-Quark}
\def\leftmark{X.-M. Xu}
 
\title{Triple-gluon and triple-quark elastic scatterings and early
thermalization} 
\authors{ 
{Xiao-Ming Xu$^1$ %
\index{Xu, X.-M.} 
}\\[2.812mm]
{\normalsize
\hspace*{-8pt}$^1$ Department of Physics,\\ Shanghai University, \\  
Baoshan, \\ 200444 Shanghai, China\\[0.2ex] 
}}
 
\abstract{Three-gluon to three-gluon scatterings lead to rapid thermalization 
of gluon matter created in central Au-Au collisions at RHIC energies. 
Thermalization of quark matter is studied from three-quark to three-quark 
scatterings.} 

 
\makeindex
\begin{document}
 
\maketitle

\section{Introduction}\label{intro}
While the energy of nucleus-nucleus collisions increases, the density of 
initially created partons gets larger and larger.   
Number density of gluon matter has been required to
 be more than 30 fm$^{-3}$ in RHIC Au-Au
collisions and may reach 140 fm$^{-3}$ in LHC Pb-Pb collisions. At such      
high gluon number densities, new physics processes begin to play an important
role and are responsible for some phenomena. Three-gluon to three-gluon 
scatterings have been shown to lead to rapid thermalization of gluon matter
created in RHIC central Au-Au collisions \cite{xmxu1}. 
Thermalization has been studied from parton cascade models 
\cite{geiger,shin,zxu} and in other model attempts 
\cite{shuryak,eskola,bhalerao,nayak,wong,baier,matinya,others}.

To study triple-gluon elastic scatterings, we need to know
how frequently the triple-gluon scatterings occur. This is shown by the ratio
of three-gluon to two-gluon scattering numbers in the next section. 
The triple-gluon and triple-quark elastic scatterings 
and transport equations are given briefly in Sections 3-5,
respectively. Summary is in the last section.

\section{Ratio of Scattering Numbers}
\label{numtonum}  

we estimate the ratio of numbers of 
scatterings occurring between two gluons or among three gluons at the moment
when gluon matter is formed. The counting of
scatterings is made for gluons in
an anisotropic momentum distribution obtained from HIJING simulation 
\cite{levai}
for initial central Au-Au collisions at $\sqrt {s_{NN}}=200$ GeV, 
\begin{equation} 
f({\bf p},t_{\rm ini})=\frac {1.07\times 10^6 (2\pi)^{1.5}} 
{\pi R_A^2 Y(\mid {\bf p} \mid/\cosh ({\rm y})+0.3)} 
{\rm e}^{-\mid {\bf p} \mid/(0.9\cosh ({\rm y}))-(\mid {\bf p} \mid  
\tanh ({\rm y}))^2/8} \bar {\theta} (Y^2-{\rm y}^2)       \label{eq1}
\end{equation} 
where 
$R_A=6.4$ fm, $t_{\rm ini}$ = 0.2 fm/$c$ and rapidity region
$\mid y \mid \leq Y=5$. $\bar {\theta} (x)$ equals 0 for $x<0$ 
or 1 for $x \ge 0$. One thousand gluons are generated from the 
distribution within the cylinder which has a radius of $R_A$ and a 
longitudinal regime of -0.2 ${\rm fm }$ $<z<0.2$ ${\rm fm}$. Then the maxima of
the numbers of the two-gluon and three-gluon scatterings are 500 and 333,
respectively. We determine a scattering of 
two gluons when the distance of the two gluons is less than a given interaction
range. If three gluons are within a sphere which center is the center-of-mass
of the three gluons and which radius equals the given interaction range, a 
scattering of the three gluons occurs. It is plotted in Fig. 1 the ratio of 
the numbers of the three-gluon scatterings to the
two-gluon scatterings at the time $t_{\rm ini}$. If the interaction range 
approaches zero from 0.1 fm, the number of the three-gluon scatterings reduces 
faster than the two-gluon scatterings;
when the interaction range is larger than 0.6 fm, both of
the scattering numbers are close to their maxima;
when it is larger than 0.1 fm, the ratio varies around 0.7.
The importance of the three-gluon scatterings is verified.

\begin{figure}[htb]
\vspace*{0.2cm}
       \insertplot{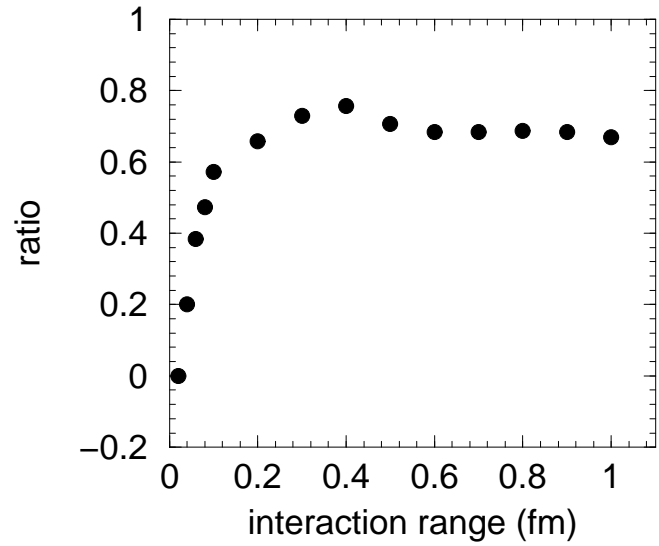}
\vspace*{-1cm}
\caption[]{Ratio of three-gluon scattering to two-gluon scattering numbers 
versus the interaction range.}
\label{num3to2}
\end{figure}

\section{Three-Gluon to Three-Gluon Scatterings}\label{gluon}  

The triple-gluon elastic scatterings involve many diagrams 
at order $\alpha_{\rm s}^4$ \cite{xmxu1} 
and only two of them are selected to be shown in Fig. 2 
for illustration. The scattering processes at tree level contain 
three-gluon and four-gluon couplings.
Three incoming gluons interact at different space-time points or at the same
space-time point. The squared four-momenta of propagators, $q_{12}^2$,
$q_{22}^2$ and $q_{23}^2$,
may tend to zero which causes Coulomb exchange divergence.
The divergence is removed with the use of a screening mass. The
triple-gluon scattering ${\rm B}_-$ 
cannot be thus identified as an iterative process 
of two successive scatterings of on-shell gluons. Such screening mass
is evaluated in the use of a formula in Ref. \cite{biro}.    

Squared amplitudes of three-gluon scattering diagrams are derived with Fortran
code in the Feynman gauge. Interference terms of different diagrams are also 
calculated. If a gluon's four-momentum is labeled as 
$p_{\rm i}=(E_{\rm i},{\bf p}_{\rm i})$ in the process
${\rm g}(p_1)+{\rm g}(p_2)+{\rm g}(p_3) \to
{\rm g}(p_4)+{\rm g}(p_5)+{\rm g}(p_6)$, 
squared amplitudes are outputed in terms of 
the following Lorentz-invariant variables, 
$s_{12}=(p_1+p_2)^2$,$s_{23}=(p_2+p_3)^2$,$s_{31}=(p_3+p_1)^2$,
$u_{15}=(p_1-p_5)^2$,$u_{16}=(p_1-p_6)^2$,
$u_{24}=(p_2-p_4)^2$,$u_{26}=(p_2-p_6)^2$,
$u_{34}=(p_3-p_4)^2$ and $u_{35}=(p_3-p_5)^2$.
In derivations, all possible exchanges of final (initial) gluons are also 
taken into account to obtain different diagrams.

\begin{figure}[htb]
\vspace*{0.2cm}
   \hspace*{-1cm}
   \begin{minipage}[htb]{8cm}
     \insertplot{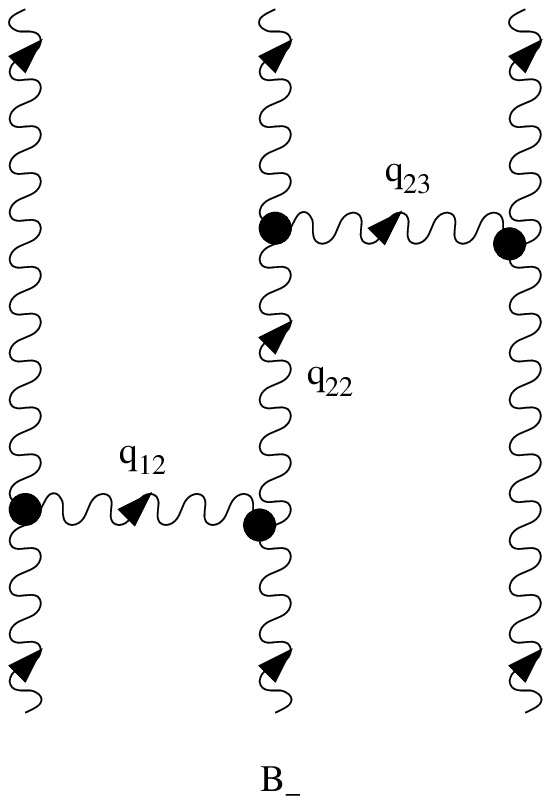}   
   \end{minipage} 
   \begin{minipage}[htb]{2cm}
     \insertplot{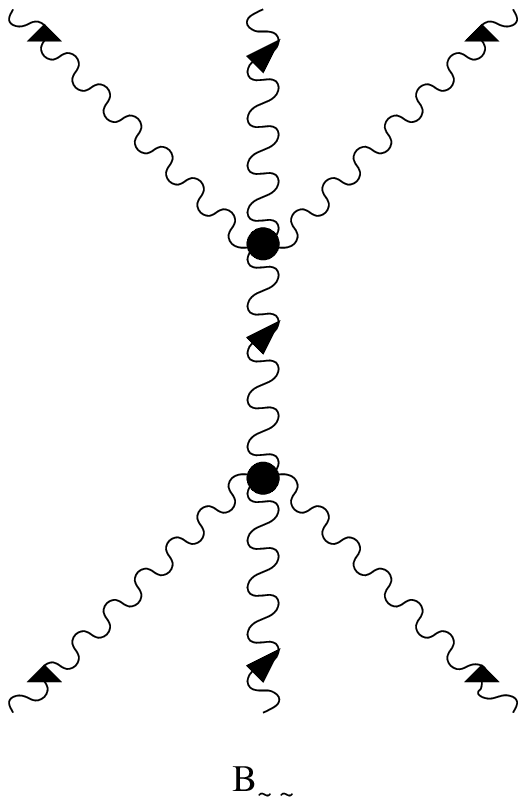}
   \end{minipage} 
\vspace*{-0.2cm}
\caption[]{Scatterings of three gluons.}
\label{ggg}
\end{figure}

\section{Three-Quark to Three-Quark Scatterings}\label{quark}  

We use quark-quark elastic scatterings \cite{cutler,combridge} and three-quark
to three-quark elastic scatterings \cite{xmxu2} to study
quark matter which is an ingredient of quark-gluon matter created in central 
Au-Au collisions at $\sqrt {s_{NN}}=200$ GeV. 
Quark matter is considered as consisting of only
up-quarks and down-quarks and the two kinds of quarks have the same 
distribution functions. 
The triple-quark elastic scattering processes are plotted in 
Fig. 3. Exchanges of final (initial) quarks generate forty-two different
diagrams . Squared amplitudes for all the diagrams are derived from Fortran 
code and expressed in terms of $s_{12}$, $s_{23}$, $s_{31}$, $u_{15}$, 
$u_{16}$, $u_{24}$, $u_{26}$, $u_{34}$ and $u_{35}$. 
The 42 diagrams contribute to the scatterings of three quarks with the
same flavor. If one quark's flavor differs from the other two, only 14 diagrams
give contributions.

\begin{figure}[htb]
\vspace*{0.2cm}
   \begin{minipage}[htb]{0.5\textwidth}
     \insertplot{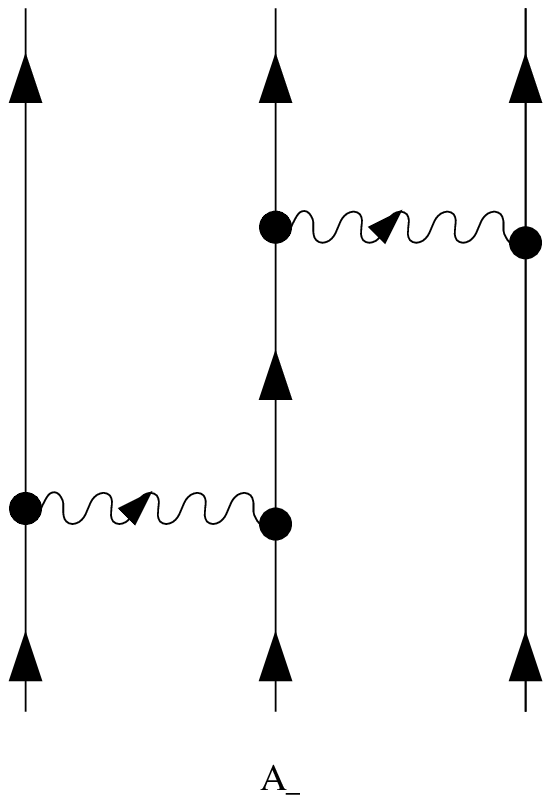}   
   \end{minipage} 
   \begin{minipage}[htb]{0.5\textwidth}
     \insertplot{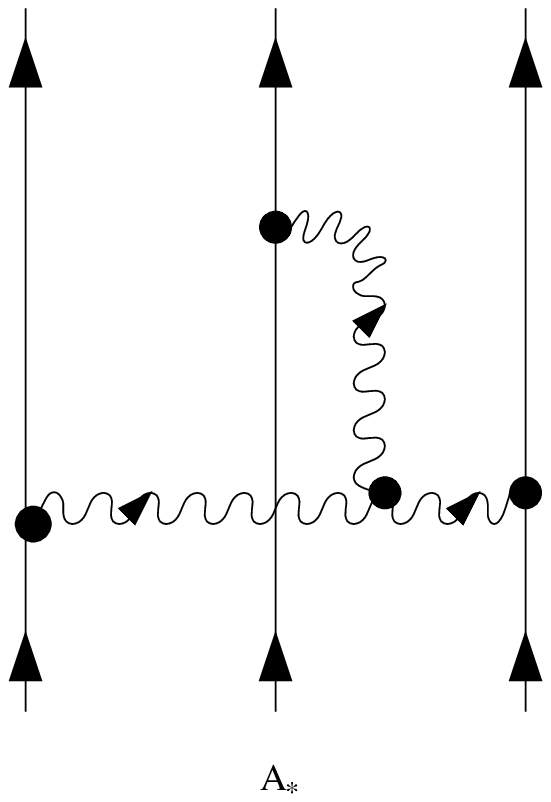}
   \end{minipage} 
\vspace*{-0.2cm}
\caption[]{Scatterings of three quarks.}
\label{qqq}
\end{figure}

\section{Transport Equations}\label{maths}

A transport equation of Boltzmann type including three-parton to three-parton
scatterings is
\begin{eqnarray}
& & 
\frac {\partial f_1}{\partial t} 
+ {\rm\bf v}_1 \cdot {\nabla}_{\bf r} f_1 
         \nonumber    \\
& &
= -\frac {\rm g}{2E_1{\rm g}_{22}} \int \frac {d^3p_2}{(2\pi)^32E_2}
\frac {d^3p_3}{(2\pi)^32E_3} \frac {d^3p_4}{(2\pi)^32E_4}  
(2\pi)^4 \delta^4(p_1+p_2-p_3-p_4)
         \nonumber    \\
& &
~~~ \times \mid {\cal M}_{2 \to 2} \mid^2
[f_1f_2(1 \pm f_3)(1 \pm f_4)-f_3f_4(1 \pm f_1)(1 \pm f_2)]
         \nonumber    \\
& &
~~~ -\frac {{\rm g}^2}{2E_1{\rm g}_{33}} 
\int \frac {d^3p_2}{(2\pi)^32E_2}
\frac {d^3p_3}{(2\pi)^32E_3} \frac {d^3p_4}{(2\pi)^32E_4}  
\frac {d^3p_5}{(2\pi)^32E_5} \frac {d^3p_6}{(2\pi)^32E_6}  
         \nonumber    \\
& &
~~~ \times (2\pi)^4 \delta^4(p_1+p_2+p_3-p_4-p_5-p_6)
\mid {\cal M}_{3 \to 3} \mid^2 
         \nonumber    \\
& &
~~~ \times [f_1f_2f_3(1 \pm f_4)(1 \pm f_5)(1 \pm f_6)
           -f_4f_5f_6(1 \pm f_1)(1 \pm f_2)(1 \pm f_3)]
         \nonumber    \\
\end{eqnarray}
where $\rm g$ is the color-spin degeneracy factor, 
the velocity for massless partons ${\rm v}_1=1$, 
${\rm g}_{22}=n^\prime_{\rm out} !$ and 
${\rm g}_{33}=n_{\rm in} ! n_{\rm out} !$ where 
$n^\prime_{\rm out}$ ($n_{\rm out}$) is the number of identical final partons
of $2 \to 2$ ($3 \to 3$) scatterings and $n_{\rm in}$ for the $3 \to 3$
scatterings is the number of 
identical initial partons except the parton in the distribution function 
$f_1$. $\mid {\cal M}_{2 \to 2} \mid^2$ and $\mid {\cal M}_{3 \to 3} \mid^2$
represent squared amplitudes of $2 \to 2$  and $3 \to 3$ parton
scatterings, respectively. The $3 \to 3$ scattering processes involve a larger
phase space than the $2 \to 2$ scattering processes. 

\begin{figure}[htb]
\vspace*{0.2cm}
       \insertplot{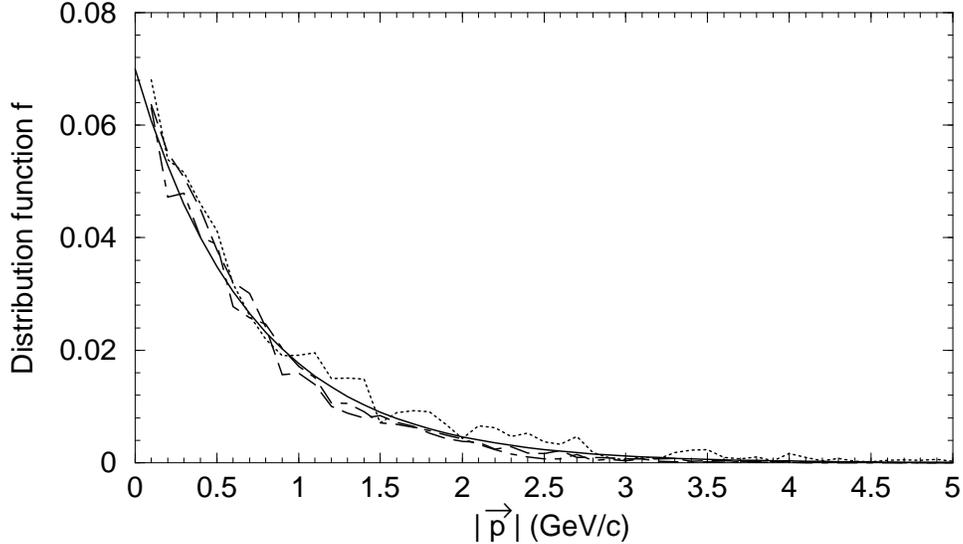}
\vspace*{-1cm}
\caption[]{Gluon distribution functions versus momentum
at $t_{\rm iso}=0.65$ fm/$c$. The
dotted, dashed and dot-dashed curves correspond to the angles 
$\theta = 0^{\rm o}, 45^{\rm o}, 90^{\rm o}$, respectively. The solid curve
represents the thermal distribution function given by Eq. (3).}
\label{thermal}
\end{figure}

Anisotropy of a parton momentum distribution as in Eq.~(\ref{eq1}) 
can be eliminated by elastic scatterings among  
partons. The transport equation is solved from the time $t_{\rm ini}$ 
when anisotropic parton matter  
is formed and until the time $t_{\rm iso}$ when local momentum isotropy is
established. 
For gluon matter, gluon distribution functions at $t_{\rm iso}=0.65$ fm/$c$
in three different directions are shown in Fig. 4 by the dotted, dashed and
dot-dashed curves. The three directions correspond to the angles
$\theta = 0^{\rm o}, 45^{\rm o}, 90^{\rm o}$ with respect to an incoming 
gold beam direction. 
The solution of the transport equation at $t_{\rm iso}$ exhibits similar
momentum dependences in different directions relative to the  
beam direction.
Such dependences can thus be fitted to the J$\rm \ddot u$ttner  
distribution which differs substantially from the anisotropic momentum 
distribution given by Eq. (\ref{eq1}), 
\begin{equation} 
f({\bf p},t_{\rm iso})=\frac {\lambda}{{\rm e}^{\mid {\bf p} \mid/T}-\lambda}
\end{equation} 
where temperature $T=0.75$ GeV, fugacity $\lambda=0.065$, 
a thermalization time of $t_{\rm iso}-t_{\rm ini}=0.45$ fm/$c$ 
for gluon matter. When quark matter evolves independently, $2 \to 2$ 
and $3 \to 3$ quark scatterings result in a thermal state with
$T=0.59$ GeV, $\lambda=0.04$ and $t_{\rm iso}-t_{\rm ini}=1.8$ fm/$c$ 
for quark matter \cite{xmxu2}. 
The thermalization times for gluon matter and quark matter are very different.

\begin{figure}[htb]
\vspace*{0.2cm}
       \insertplot{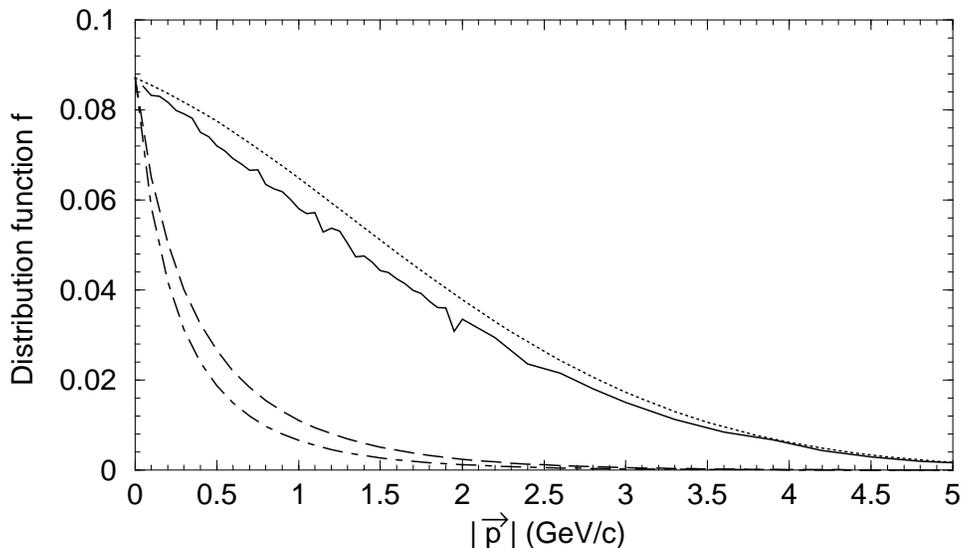}
\vspace*{-1cm}
\caption[]{Gluon distribution functions at $t_{\rm ini}=0.2$ fm/$c$ are 
shown by the
dotted, dashed and dot-dashed curves which correspond to the angles 
$\theta = 0^{\rm o}, 45^{\rm o}, 90^{\rm o}$, respectively. The $2 \to 2$
gluon scatterings result in the solid curve
for the gluon distribution function at $\theta = 0^{\rm o}$
at $t_{\rm iso}=0.65$ fm/$c$.}
\label{gg}
\end{figure}

To definitely realize a role of the $3 \to 3$ gluon scatterings, the gluon
distribution function at the time $t_{\rm iso}=0.65$ fm/$c$ resulted only from 
the $2 \to 2$ gluon scatterings is shown as a solid curve in Fig. 5. 
The small discrepancy of the solid and the 
dotted curves exhibits a small variation of the gluon distribution function at
the angle $\theta=0^{\rm o}$.  The gluon distribution functions at the two
angles $\theta=45^{\rm o}, 90^{\rm o}$ at $t_{\rm iso}=0.65$ fm/c are very
close to the dashed and dot-dashed curves and are thus not plotted.
The variation of gluon distribution 
function resulted from the $2 \to 2$ gluon
scatterings is small in the forward
direction and even negligible away from this direction. Therefore, in the
thermalization driven by both the $2 \to 2$ and $3 \to 3$ gluon
scatterings, the gluon distribution variation from the anisotropic form to
the thermal state is mainly caused by the 
$3 \to 3$ gluon scatterings.

\section{Conclusions}\label{concl}

We have studied thermalization of gluon matter and quark matter 
with the three-parton to three-parton elastic scatterings.
The three-gluon elastic scattering processes give a considerably larger 
variation of the gluon 
distribution function than the two-gluon elastic scattering 
processes while the gluon number density is high. The triple-gluon scatterings
are important at RHIC energies and yield the effect of rapid thermalization.
The triple-quark scatterings give a variation of quark distribution function
comparable to the one that resulted from the quark-quark scatterings. Quark
matter itself cannot thermalize rapidly at RHIC energies.

\section*{Acknowledgment(s)}
This work was supported in part by National Natural Science Foundation of China
under Grant No. 10135030 and No. 10510201129, 
in part by Shanghai Education Committee Research  
Fund No. 04AB04 and in part by the CAS Knowledge Innovation Project 
No. KJCX2-SW-N02.

\vfill\eject

\begin{thebibliography}{99}  
  
\bibitem{xmxu1}X.-M. Xu, Y. Sun, A.-Q. Chen and L. Zheng, {\it Nucl. Phys.}
 {\bf A744} (2004) 347.
 
\bibitem{geiger}K. Geiger, {\it Phys. Rev.}  {\bf D46} (1992) 4965; 
{\it Phys. Rev.} {\bf D46} (1992) 4986.
 
\bibitem{shin}G.R. Shin and B. M$\rm \ddot u$ller, {\it J. Phys.}  
{\bf G29} (2003) 2485.

\bibitem{zxu}Z. Xu and C. Greiner, {\it Phys. Rev.} {\bf C71} (2005) 064901.

\bibitem{shuryak}E. Shuryak, {\it Phys. Rev. Lett.} {\bf 68} (1992) 3270.

\bibitem{eskola}K.J. Eskola and M. Gyulassy, {\it Phys. Rev.} {\bf C47} 
(1993) 2329.

\bibitem{bhalerao}R.S. Bhalerao and G.C. Nayak, {\it Phys. Rev.} {\bf C61} 
(2000) 054907.

\bibitem{nayak}G.C. Nayak, A. Dumitru, L. McLerran and W. Greiner, 
{\it Nucl. Phys.} {\bf A687} (2001) 457.

\bibitem{wong}S.M. Wong, {\it Phys. Rev.} {\bf C54} (1996) 2588.

\bibitem{baier}R. Baier, A.H. Mueller, D. Schiff and D.T. Son, 
{\it Phys. Lett.} {\bf B502} (2001) 51.

\bibitem{matinya}S.G. Matinyan, B. M$\rm \ddot u$ller and D.H. Rischke, 
{\it Phys. Rev.} {\bf C57} (1998) 1927.

\bibitem{others}See talks in this issue.

\bibitem{levai}P. L$\rm \acute e$vai, B. M$\rm \ddot u$ller, X.-N. Wang, 
{\it Phys. Rev.} {\bf C51} (1995) 3326.

\bibitem{biro}T.S. Bir$\rm \acute o$, B. M$\rm \ddot u$ller,
X.-N. Wang, {\it Phys. Lett.} {\bf B283} (1992) 171.

\bibitem{cutler}R. Cutler, D. Sivers, {\it Phys. Rev.} {\bf D17} (1978) 196.

\bibitem{combridge}B.L. Combridge, J. Kripfganz, J. Ranft, {\it Phys. Lett.} 
{\bf B70} (1977) 234.

\bibitem{xmxu2}X.-M. Xu, R. Peng, H.J. Weber,   
{\it Phys. Lett.} {\bf B629} (2005) 68.





\end{thebibliography}
\end{document}